\documentclass[10pt, letterpaper,twocolumn]{article}

\usepackage[margin=2.54cm]{geometry}
\usepackage{lineno}
\usepackage{amsmath,amssymb}
\usepackage{graphicx}
\usepackage[round,comma,authoryear]{natbib}
\bibpunct[]{(}{)}{,}{a}{}{,}
\usepackage{float}
\usepackage{setspace}
\usepackage{comment}
\usepackage{booktabs}
\usepackage{lscape}
\usepackage{titlesec}
\usepackage{rotating}
\usepackage{nidanfloat}

\titleformat*{\section}{\normalsize\bfseries}
\titleformat*{\subsection}{\normalsize\bfseries}

\title{
The forecasting of menstruation based on a 
state-space \\ modeling of basal body temperature time series
}

\begin{small}
\author{
  {\small Keiichi Fukaya,$^{1,4}$ Ai Kawamori,$^{1}$ Yutaka Osada,$^{2}$
    Masumi Kitazawa,$^{3}$ and Makio Ishiguro$^{1}$} \\\\
  {\footnotesize \it $^{1}$The Institute of Statistical Mathematics, 10-3 Midoricho, Tachikawa, Tokyo 190-8562 Japan} \\
  {\footnotesize \it $^{2}$Research Institute for Humanity and Nature, 457-4 Motoyama, Kamigamo, Kita-ku, Kyoto, 603-8047 Japan} \\
  {\footnotesize \it $^{3}$QOL Corporation, 3-15-1, Tokida, Ueda, Nagano 386-8567 Japan} \\
  {\footnotesize $^{4}$kfukaya@ism.ac.jp}
}
\end{small}

\begin{document}

\date{ }

\twocolumn[
  \begin{@twocolumnfalse}
    \maketitle

    \begin{abstract}
      Women's basal body temperature (BBT) follows a periodic pattern that
      is associated with the events in their menstrual cycle.
      Although daily BBT time series contain potentially useful information for
      estimating the underlying menstrual phase and for predicting the length of
      current menstrual cycle, few models have been constructed for BBT time series.
      Here, we propose a state-space model that includes menstrual phase
      as a latent state variable to explain fluctuations in BBT
      and menstrual cycle length.
      Conditional distributions for the menstrual phase were obtained by using 
      sequential Bayesian filtering techniques.
      A predictive distribution for the upcoming onset of menstruation was then derived
      based on the conditional distributions and the model,
      leading to a novel statistical framework 
      that provided a sequentially updated prediction of the day of onset of menstruation.
      We applied this framework to a real dataset comprising women's self-reported BBT
      and days of menstruation,
      comparing the prediction accuracy of our proposed method 
      with that of conventional calendar calculation.
      We found that our proposed method provided a better prediction
      of the day of onset of menstruation.
      Potential extensions of this framework
      may provide the basis of modeling and predicting other events that are
      associated with the menstrual cycle.\\

    {\noindent \em Key words:} 
    {\em Basal body temperature, Menstrual cycle length, Periodic phenomena, 
      Sequential prediction, State-space model, Time series analysis.}\\\\
    \end{abstract}
  \end{@twocolumnfalse}
  ]

\section{Introduction}

The menstrual cycle is the periodic changes that occur in the
female reproductive system that make pregnancy possible.
Throughout the menstrual cycle, basal body temperature (BBT) also follows a periodic pattern.
The menstrual cycle consists of two phases, the follicular phase followed by
the luteal phase, with ovulation occurring at the transition between the two phases. 
During the follicular phase, BBT is relatively low with the nadir occurring
within 1 to 2 days of a surge in luteinizing hormone that triggers ovulation.
After the nadir, the cycle enters the luteal phase and 
BBT rises by $0.3$ to $0.5\ {}^\circ\mathrm{C}$  \citep{Barron2005}.

Considerable attention has been paid to the development of methods
to predict the days of ovulation and the day of onset of menstruation;
currently available methods include urinary and plasma hormone analyses,
ultrasound monitoring of follicular growth, and monitoring 
of changes in the cervical mucus or BBT. 
Monitoring the change in BBT is straightforward because it requires neither
expensive instruments nor medical expertise.
However, of the currently available methods, BBT measurement is 
the least reliable because it has an inherently large day-to-day variability \citep{Barron2005};
therefore, statistical analyses are required to improve 
predictions of events associated with the menstrual cycle based on BBT.

Many statistical models of the menstrual cycle have been proposed,
with the majority explaining the marginal distribution of menstrual cycle length
which is characterized by a long right tail.
To explain within-individual heterogeneity, which is an important source
of variation in menstrual cycle length, \cite{Harlow1991} made the biological assumption
that a single menstrual cycle is composed of a period of ``waiting''
followed by the ovarian cycle.
They classified menstrual cycles into standard, normally distributed
cycles that contained no waiting time and into nonstandard cycles that
contained waiting time.
\cite{Guo2006} extended this idea and proposed a mixture model consisting of a
normal distribution and a shifted Weibull distribution to explain the right-tailed distribution.
By accommodating covariates, they also investigated the effect of 
an individual's age on the moments of the cycle length distribution.
\cite{Huang2014} attempted to model the changes in the mean and variance of 
menstrual cycle length that occur during the approach of menopause. 
By using a change-point model, they found changes in the annual rate of 
change of the mean and variance of menstrual cycle length which
indicate the start of early and late menopausal transition.
In contrast,  \cite{Bortot2010} focused on the dynamic aspect of
menstrual cycle length over time. By using a state-space modeling approach, 
they derived a predictive distribution of menstrual cycle length
that was conditional on past time series.
By integrating a fecundability model into their time series model, 
they also developed a framework that estimates the probability of conception 
that is conditional on the within-cycle intercourse behavior.
A related joint modeling of menstrual cycle length and fecundity has been
attained more recently \citep{Lum2015}.

Although daily fluctuations in BBT are associated with the events of the menstrual cycle,
and therefore BBT time series analyses likely provide information regarding 
the length of the current menstrual cycle,
there have been no previous studies modeling BBT data to predict menstrual cycle length.
Thus, in the present study we developed a statistical framework that provides
a predictive distribution of menstrual cycle length (which, by extension, is 
a predictive distribution of the next day of onset of menstruation) that is 
sequentially updated with daily BBT data.
We used a state-space model that includes a latent phase state variable
to explain daily fluctuations in BBT and to derive a predictive distribution
of menstrual cycle length that is dependent on the current phase state.

This paper is organized as follows: 
In Section \ref{sect:data}, we briefly describe the data required for
the proposed method and how the test dataset was obtained.
In Section \ref{sect:model}, we provide 
the formulation of the proposed state-space model for menstrual cycle 
and give an overview of the filtering algorithms that we use to estimate the
conditional distributions of the latent phase variable given the model and dataset.
In the same section, we also describe 
the predictive probability distribution for the next day of onset of menstruation
that was derived from the proposed model and the filtering distribution for the menstrual phase.
In Section \ref{sect:appl}, we apply the proposed framework to a real dataset
and compare the accuracy of the point prediction of the next day of onset of menstruation 
between the proposed method and the conventional method of calendar calculation.
In Section \ref{sect:disc}, we discuss the practical utility of the proposed method 
with respect to the management of women's health and examine the potential challenges
and prospects for the proposed framework.

\section{Data}
\label{sect:data}

We assumed that for a given female subject a dataset containing
a daily BBT time series and days of onset of menstruation was available.
The BBT could be measured with any device (e.g., a conventional thermometer
or a wearable sensor), although different model parameters may be adequate
for different measurement devices.
In the application of the proposed framework described in Section \ref{sect:appl},
we used real BBT time series and menstruation onset data that was
collected via a website called {\it Ran's story} (QOL Corporation, Ueda, Japan),
which is a website that allows registered users to upload their self-reported
daily BBT and days of menstruation onset to QOL Corporation's data servers.
At the time of registering to use the service, 
all users of {\it Ran's story} agree to the use of their data
for academic research.
Although no data regarding the ethnic characteristics of the users were available,
it is assumed that the majority, if not all, the users were ethnically Japanese
because {\it Ran's story} is provided only in the Japanese language.

\section{Model description and inferences}
\label{sect:model}

\subsection{State-space model of the menstrual cycle}
\label{sect:ssm}

Here we develop a state-space model for a time series of observed BBT,
$y_t$, and an indicator of the onset of menstruation, $z_t$, obtained for a 
subject for days $t=1,\dots,T$.
By $z_t = 1$, we denote that menstruation started on day $t$,
whereas $z_t = 0$ indicates that day $t$ is not the first day of menstruation.
We denote the BBT time series and menstruation data obtained until time $t$ as
$Y_t = (y_1, \dots, y_t)$ and $Z_t = (z_1, \dots, z_t)$, respectively.

We considered the phase of the menstrual cycle,
$\theta_t \in \mathbb{R}~(t = 0, 1, \dots, T)$,
to be a latent state variable.
We let $\epsilon_t$ as the daily advance of the phase
and assume that it is a positive random variable that follows a gamma distribution
with shape parameter $\alpha$ and rate parameter $\beta$,
which leads to the following system model:
%
\begin{align}
  & \theta_t = \theta_{t-1} + \epsilon_t \\
  & \epsilon_t \sim \textrm{Gamma}(\alpha, \beta).
\end{align}
Under this assumption, the conditional distribution of $\theta_t$, given $\theta_{t-1}$,
is a gamma distribution with a probability density function:
%
\begin{align}
  &p(\theta_t \mid \theta_{t-1}) 
  = \textrm{Gamma}(\alpha, \beta) \notag \\
  &= \frac{\beta^\alpha}{\Gamma(\alpha)}
    (\theta_t - \theta_{t-1})^{\alpha - 1}
    \exp \left\{-\beta (\theta_t - \theta_{t-1}) \right\}.
\end{align}

It is assumed that the distribution for the observed BBT $y_t$ 
is conditional on the phase $\theta_t$.
Since periodic oscillation throughout each phase is expected for BBT,
a finite trigonometric series is used to model the average BBT.
Assuming a Gaussian observation error, the observation model for BBT is expressed as
%
\begin{align}
  y_t &= a + \sum_{m=1}^M 
      (b_m\cos2m\pi\theta_t + c_m\sin2m\pi\theta_t) + e_t \hspace{2em} \\
  e_t &\sim \textrm{Normal}(0, \sigma^2),
\end{align}
\noindent{where} $M$ is the maximum order of the series.
Conditional on $\theta_t$, $y_t$ then follows a normal distribution with a
probability density function:
%
\begin{align}
  p(y_t \mid \theta_t) &= \textrm{Normal}\left\{\mu(\theta_t), \sigma^2 \right\} \notag \\
  & = \frac{1}{\sqrt{2\pi\sigma^2}}
  \exp \left[ -\frac{\left\{y_t - \mu(\theta_t)\right\}^2}{2\sigma^2} \right],
\end{align}
\noindent{where} $\mu(\theta_t) = a + \sum_{m=1}^M 
      (b_m\cos2m\pi\theta_t + c_m\sin2m\pi\theta_t)$.
By this definition, $\mu(\theta_t)$ is periodic in terms of $\theta_t$ with a period of 1.

For the onset of menstruation, we assume that 
menstruation starts when $\theta_t$ ``steps over'' the smallest following integer.
This is represented as follows:
%
\begin{align}
  z_t &= 0 \hspace{2em} \textrm{when} \hspace{2em}
    \lfloor \theta_t \rfloor = \lfloor \theta_{t-1} \rfloor \label{eqn:obs_mens1}\\
  &= 1 \hspace{2em} \textrm{when} \hspace{2em}
    \lfloor \theta_t \rfloor > \lfloor \theta_{t-1} \rfloor, \label{eqn:obs_mens2}
\end{align}
\noindent{where} $\lfloor x \rfloor$ is the floor function that returns
the largest previous integer for $x$.
Writing this deterministic allocation in a probabilistic manner, 
which is conditional on $(\theta_t, \theta_{t-1})$,
$z_t$ follows a Bernoulli distribution:
%
\begin{align}
  &p(z_t \mid \theta_t, \theta_{t-1}) \notag \\
  & = (1 - z_t)\left\{I(\lfloor \theta_t \rfloor = \lfloor \theta_{t-1} \rfloor)\right\} + 
    z_t\left\{I(\lfloor \theta_t \rfloor > \lfloor \theta_{t-1} \rfloor)\right\},
\end{align}

\noindent{where} $I(x)$ is the indicator function that returns 1 when $x$ is 
true or 0 otherwise.

Let $\boldsymbol{\xi} = (\alpha, \beta, \sigma, a, b_1, \dots, b_M, c_1, \dots c_M)$
be a vector of the parameters of this model.
Given a time series of BBT, $Y_T = (y_1, \dots, y_T)$, an indicator of menstruation,
$Z_T = (z_1, \dots, z_T)$, and a distribution specified for 
initial states, $p(\theta_1, \theta_0)$,
these parameters can be estimated by using the maximum likelihood method.
The log-likelihood of this model is expressed as
%
\begin{align}
  &l(\boldsymbol{\xi}; Y_T, Z_T) \notag \\
  &=\log p(y_1,z_1 \mid \boldsymbol{\xi})+
  \sum_{t=2}^T\log p(y_t,z_t \mid Y_{t-1}, Z_{t-1}, \boldsymbol{\xi}),
\end{align}
\noindent{where}
%
\begin{align}
  &\log p(y_1,z_1 \mid \boldsymbol{\xi}) \notag \\
  &= \log \int\int p(y_1 \mid \theta_1)p(z_1 \mid \theta_1, \theta_0)p(\theta_1, \theta_0) d\theta_1 d\theta_0,  \label{eqn:loglik1}
\end{align}
\noindent{and} for $t = 2, \dots, T$,
%
\begin{align}
  &\log p(y_t,z_t \mid Y_{t-1}, Z_{t-1}, \boldsymbol{\xi}) \notag \\
  &= \log \int\int p(y_t \mid \theta_t)p(z_t \mid \theta_t, \theta_{t-1}) \notag \\
  &\hspace{5em}\times p(\theta_t, \theta_{t-1} \mid Y_{t-1}, Z_{t-1}) d\theta_t d\theta_{t-1},  \label{eqn:loglikt}
\end{align}
which can be sequentially obtained by using the Bayesian filtering technique described below.
Note that although $p(y_t \mid \theta_t) (t\geq1)$ and 
$p(\theta_t, \theta_{t-1} \mid Y_{t-1}, Z_{t-1}) (t\geq2)$ depend on $\boldsymbol{\xi}$,
this dependence is not explicitly described for notational simplicity.

\subsection{State estimation and calculation of log-likelihood by using the non-Gaussian filter}
\label{sect:ngf}

Given the state-space model of menstrual cycle described above and its parameters and data,
the conditional distribution of an unobserved menstrual phase can be obtained by using 
recursive formulae for the state estimation problem,
which are referred to as the Bayesian filtering and smoothing equations \citep{Sarkka2013}.
We describe three versions of this sequential procedure, that is, prediction,
filtering, and smoothing, for the state-space model described above.

Let $p(\theta_t, \theta_{t-1} \mid Y_t, Z_t)$
be the joint distribution for the phase at successive time points $t-1$ and $t$, 
which is conditional on the observations obtained by time $t$.
This conditional distribution accommodates all the data obtained by time $t$ and
is called the filtering distribution.
Similarly, the joint distribution for the phase of successive time points $t$ and $t-1$
conditional on the observations obtained by time $t-1$,
$p(\theta_t, \theta_{t-1} \mid Y_{t-1}, Z_{t-1})$,
is referred to as the one-step-ahead predictive distribution.
For $t = 1,\dots,T$ these distributions are obtained 
by sequentially applying the following recursive formulae:

\vspace{-.2em}
{\footnotesize
\begin{align}
  &p(\theta_t, \theta_{t-1} \mid Y_{t-1}, Z_{t-1}) \notag \\
  & = p(\theta_t \mid \theta_{t-1})p(\theta_{t-1} \mid Y_{t-1},Z_{t-1}) \notag \\ 
  & = p(\theta_t \mid \theta_{t-1}) \int p(\theta_{t-1}, \theta_{t-2} \mid Y_{t-1},Z_{t-1}) d\theta_{t-2} \label{eqn:predictor}\\
  &p(\theta_t, \theta_{t-1} \mid Y_t, Z_t) \notag \\
  & = \frac{p(y_t, z_t \mid \theta_t, \theta_{t-1})p(\theta_t, \theta_{t-1} \mid Y_{t-1}, Z_{t-1})}{\int\int p(y_t, z_t \mid \theta_t, \theta_{t-1})p(\theta_t, \theta_{t-1} \mid Y_{t-1}, Z_{t-1}) d\theta_t d\theta_{t-1}}\notag \\
  & = \frac{p(y_t \mid \theta_t)p(z_t \mid \theta_t, \theta_{t-1})p(\theta_t, \theta_{t-1} \mid Y_{t-1}, Z_{t-1})}{\int\int p(y_t \mid \theta_t)p(z_t \mid \theta_t, \theta_{t-1})p(\theta_t, \theta_{t-1} \mid Y_{t-1}, Z_{t-1}) d\theta_t d\theta_{t-1}} \label{eqn:filter} 
\end{align}
}

\noindent{where} for $t = 1$ we set $p(\theta_1, \theta_0 \mid Y_0, Z_0)$
as $p(\theta_1, \theta_0)$, which is the specified initial distribution for the phase.
Equations (\ref{eqn:predictor}) and (\ref{eqn:filter})
are the prediction and filtering equation, respectively.
Note that the denominator in Equation (\ref{eqn:filter}) is the likelihood
for data at time $t$ (see Equations \ref{eqn:loglik1} and \ref{eqn:loglikt}).
Hence, the log-likelihood of the state-space model is obtained
through application of the Bayesian filtering procedure.

The joint distribution for the phase, which is conditional on the entire set of observations,
$p(\theta_t, \theta_{t-1} \mid Y_T, Z_T)$, 
is referred to as the (fixed-interval type) smoothed distribution.
With the filtering and the one-step-ahead predictive distributions, 
the smoothed distribution is obtained by recursively using the 
following smoothing formula:

\vspace{-.2em}
{\footnotesize
\begin{align}
  &p(\theta_t, \theta_{t-1} \mid Y_T, Z_T) \notag \\
    & = p(\theta_t, \theta_{t-1} \mid Y_t, Z_t)
      \int \frac{p(\theta_{t+1}, \theta_t \mid Y_T, Z_T)
        p(\theta_{t+1} \mid \theta_t)}
        {p(\theta_{t+1}, \theta_t \mid Y_t, Z_t)} d\theta_{t+1} \notag \\
        & = \frac{p(\theta_t, \theta_{t-1} \mid Y_t, Z_t) \int p(\theta_{t+1}, \theta_t \mid Y_T, Z_T) d\theta_{t+1}}
      {p(\theta_t \mid Y_t, Z_t)} \notag \\
        & = \frac{p(\theta_t, \theta_{t-1} \mid Y_t, Z_t) \int p(\theta_{t+1}, \theta_t \mid Y_T, Z_T) d\theta_{t+1}}
        {\int p(\theta_t, \theta_{t-1} \mid Y_t, Z_t) d\theta_{t-1}} \label{eqn:smoother}.
\end{align}
}

In general, these recursive formulae are not analytically tractable for 
non-linear, non-Gaussian, state-space models.
However, the conditional distributions, as well as the log-likelihood
of the state-space model, can still be approximated
by using filtering algorithms for general state-space models.
We use Kitagawa's non-Gaussian filter \citep{Kitagawa1987}
where the continuous state space is discretized into
equally spaced grid points at which the probability density is evaluated.
We describe the numerical procedure to obtain these conditional
distributions by using the non-Gaussian filter in Appendix \ref{sect:appa}.

\subsection{Predictive distribution for the day of onset of menstruation}
\label{sect:pred}

A predictive distribution for the day of onset of menstruation is derived
from the assumptions of the model and the filtering distribution of the state.

We denote the distribution function and the probability density function of
the gamma distribution with shape parameter $s$ and rate parameter $r$
as $G(\hspace{.1em}\cdot\hspace{.1em} ; s, r)$ and
$g(\hspace{.1em}\cdot\hspace{.1em} ; s, r)$, respectively.
Let the accumulated advance of the phase be denoted by $\Delta_k(t)$,
which is then calculated as
$\Delta_k(t)=\sum_{r=t+1}^{t+k}\epsilon_r, k=1,2,\cdots$.
We consider the conditional probability that 
the next menstruation has occurred before day $k + t$
given the phase state $\theta_t$.
We denote this conditional probability as
$F(k\mid\theta_t) = \textrm{Pr}\left\{\Delta_k(t) > \lceil \theta_t \rceil - \theta_t\right\}$,
where $\lceil x \rceil$ is the ceiling function that returns the
smallest following integer for $x$.
Therefore, $F(k\mid\theta_t)$ represents the conditional distribution function 
for the onset of menstruation. Under the assumption of the 
state-space model described above, $F(k\mid\theta_t)$ is given as
%
\begin{align}
  F(k \mid \theta_t) &= \int_{\lceil \theta_t \rceil-\theta_t}^{\infty} 
    g(x; k\alpha, \beta) dx \notag \\
  &= 1 - G(\lceil \theta_t \rceil-\theta_t; k\alpha, \beta).
\end{align}
The conditional probability function for the day of onset of menstruation,
denoted as $f(k \mid \theta_t)$, is then given as 
%
\begin{align}
  &f(k \mid \theta_t) \notag \\
  &= F(k \mid \theta_t) - F(k-1 \mid \theta_t) \notag \\
  &= \left\{ 1 - G(\lceil \theta_t \rceil -\theta_t; k\alpha, \beta) \right\} \notag \\
  &\hspace{4em}- \left[ 1 - G\left\{\lceil \theta_t \rceil-\theta_t; (k-1)\alpha, \beta\right\} \right] \notag \\
  &=  G\left\{\lceil \theta_t \rceil-\theta_t; (k-1)\alpha, \beta\right\} 
      - G(\lceil \theta_t \rceil-\theta_t; k\alpha, \beta),
\end{align}
\noindent{where} we set $F(0 \mid \theta_t)=0$.

The marginal distribution for the day of onset of menstruation,
denoted as $h(k \mid Y_t, Z_t)$,
can also be obtained with the marginal filtering distribution for the phase state,
$p(\theta_t \mid Y_t, Z_t)$. It is given as
%
\begin{equation}
  h(k \mid Y_t, Z_t) = \int f(k \mid \theta_t)p(\theta_t \mid Y_t, Z_t) d\theta_t.
\end{equation}
Note that this distribution is conditional on the data obtained by time $t$
and therefore provides a predictive distribution for menstruation day
that accommodates all information available.
The point prediction for menstruation day can also be obtained from $h(k \mid Y_t, Z_t)$
and one of the natural choice for it would be the $k$ that gives the highest probability,
$\max h(k \mid Y_t, Z_t)$.

\begin{table*}[tb]
  \centering
  \caption{Summary of self-reported menstrual cycle data obtained from 10 subjects. }
  {\tiny
  \begin{tabular}{lcccccccccc} \toprule
    & \multicolumn{10}{c}{Subject} \\
     & 1 & 2 & 3 & 4 & 5 & 6 & 7 & 8 & 9 & 10 \\
    \midrule
      All data  \\
      \hspace{2em}No. of consecutive cycles              & 47 & 46 & 49 & 55 & 53 & 58 & 57 & 53 & 46 & 45  \\
      \hspace{2em}Range of cycle length                  & [28, 59] & [23, 45] & [19, 61] & [26, 59] & [18, 47] & [25, 49] & [25, 36] & [26, 56] & [30, 51] & [27, 48]  \\
      \hspace{2em}Mean of cycle length                   &   39.4 &   33.2 &   32.8 &   31.2 &   27.2 &   29.7 &   30.1 &   31.7 &   34.1 &   33.4 \\
      \hspace{2em}Median of cycle length                 &   38 &   33 &   32.5 &   30.5 &   26 &   29 &   30 &   31 &   32 &   33 \\
      \hspace{2em}SD of cycle length                     &    6.7 &    4.1 &    5.6 &    4.7 &    4.9 &    3.9 &    2.6 &    5.2 &    4.7 &    4.6 \\
      \hspace{2em}Initial age                            &   29.6 &   24.9 &   23.2 &   26.3 &   30.5 &   33.2 &   34.9 &   29.9 &   33.9 &   31.4 \\
      \hspace{2em}Final age                              &   34.5 &   29.0 &   27.6 &   30.9 &   34.4 &   37.8 &   39.5 &   34.4 &   38.1 &   35.4 \\
      \hspace{2em}Length of time series                  & 1812 & 1495 & 1576 & 1687 & 1418 & 1693 & 1687 & 1647 & 1534 & 1470 \\
      \hspace{2em}No. of missing observations             &   80 &    0 &   96 &   31 &   43 &   33 &   21 &   85 &   38 &   60 \\
      Data for parameter estimation \\
      \hspace{2em}No. of consecutive cycles  & 29       & 29       & 29       & 29       & 29       & 29       & 29       & 29       & 29       & 29   \\
      \hspace{2em}Range of cycle length      & [31, 59] & [23, 43] & [19, 39] & [26, 59] & [22, 37] & [25, 49] & [27, 36] & [26, 56] & [30, 51] & [27, 42]  \\
      \hspace{2em}Mean of cycle length       & 41.3     & 33.7     & 31.5     & 30.6     & 26.7     & 30.4     & 30.9     & 33.0     & 34.7     & 32.1 \\
      \hspace{2em}Median of cycle length     & 41       & 34       & 32       & 29       & 26       & 30       & 31       & 32       & 32       & 31   \\
      \hspace{2em}SD of cycle length         & 7.2      & 4.0      & 3.9      & 5.8      & 3.3      & 4.8      & 2.5      & 5.9      & 5.8      & 3.5  \\
      \hspace{2em}Length of time series      & 1199     & 978      & 914      & 888      & 776      & 882      & 897      & 958      & 1007     & 933  \\
      \hspace{2em}No. of missing observations & 24       & 0        & 52       & 21       & 20       & 22       & 8        & 5        & 19       & 34   \\
      Data for predictive accuracy estimation \\
      \hspace{2em}No. of consecutive cycles  & 18       & 17       & 20       & 26       & 24       & 29       & 28       & 24       & 17       & 16   \\
      \hspace{2em}Range of cycle length      & [28, 44] & [28, 45] & [26, 61] & [28, 39] & [18, 47] & [26, 36] & [25, 36] & [26, 39] & [31, 35] & [27, 48] \\
      \hspace{2em}Mean of cycle length       & 36.1     & 32.3     & 34.8     & 32.0     & 27.9     & 29.0     & 29.3     & 30.0     & 32.9     & 35.8 \\
      \hspace{2em}Median of cycle length     & 35       & 32       & 33       & 32       & 27       & 28       & 30       & 29       & 33       & 35   \\
      \hspace{2em}SD of cycle length         & 4.2      & 4.2      & 7.2      & 2.7      & 6.5      & 2.5      & 2.5      & 3.6      & 1.4      & 5.6  \\
      \hspace{2em}Length of time series      & 613      & 517      & 662      & 799      & 642      & 811      & 790      & 689      & 527      & 537  \\
      \hspace{2em}No. of missing observations & 56       & 0        & 44       & 10       & 23       & 11       & 13       & 80       & 19       & 26   \\
    \bottomrule
  \end{tabular}
  }
  \label{table:data}
\end{table*}

We describe the non-Gaussian filter-based numerical procedure
we used to obtain these predictive distributions in Appendix \ref{sect:appa}.

\section{Application}
\label{sect:appl}

We used the BBT time series and menstruation onset data provided by 10 
users of the {\it Ran's story} website (QOL Corporation, Ueda, Japan).
Data were collected through the course of 44 to 57 consecutive menstrual cycles
(see Table \ref{table:data} for a summary of the data).
Data of each subject's first 29 consecutive menstrual cycles were used to fit the
state-space model described above and to obtain maximum likelihood estimates of the parameters.
To determine the order of the trigonometric series,
models with $M=1,2,\dots,12$ (referring to models M1, M2, \dots, and M12, respectively) 
were fitted and then compared based on the Akaike information criterion (AIC)
for each subject.
The best AIC model and the remaining menstrual cycle data
(i.e., the data that were not used for the parameter estimations),
were then used to evaluate the accuracy of the prediction of
the next day of onset of menstruation based on the root mean square error (RMSE) 
and the mean absolute error (MAE) for each subject.
We compared the prediction accuracy 
between the sequential prediction and the conventional, fixed prediction, 
where the former was obtained by using the proposed method and the latter
was obtained as the day after a fixed number of days from the onset of preceding menstruation.

\begin{figure*}[htb]
\begin{center}
\includegraphics[bb=0 0 864 468,width=6in]{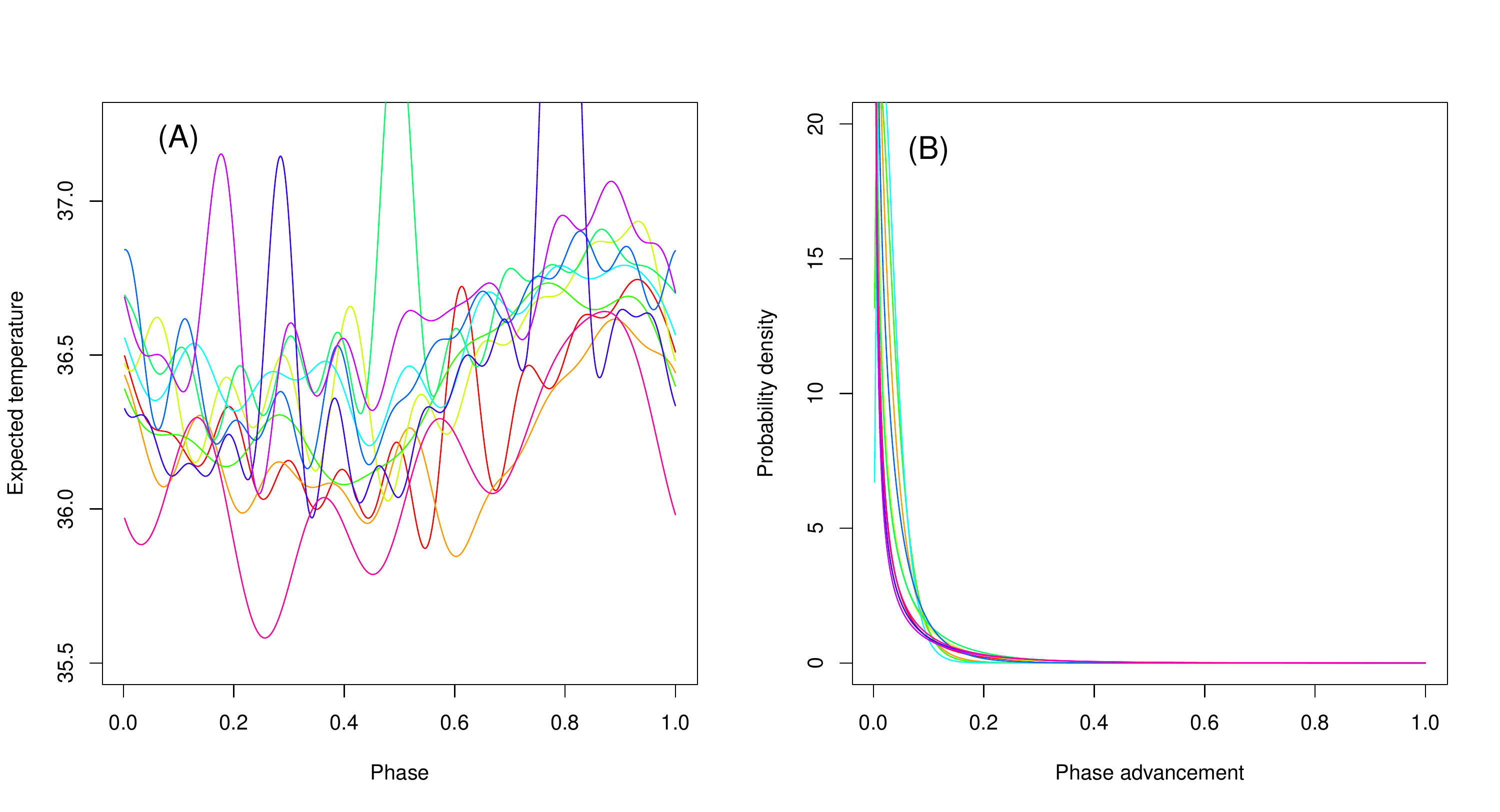}
\end{center}
\caption{
  Model components for each subject.
  Each color represents a different subject.
  Lines show the best AIC model associated with the
  maximum likelihood estimates for each subject.
  (A) Relationship between expected temperature and menstrual phase.
  The $x$-axis represents $\theta_t \mod 1$.
  (B) Probability density distribution for the advancement per day of
  the menstrual phase.
}
\label{fig:all}
\end{figure*}

\begin{sidewaysfigure*}[p]
\begin{center}
\includegraphics[bb=0 0 1296 468,width=9.1in]{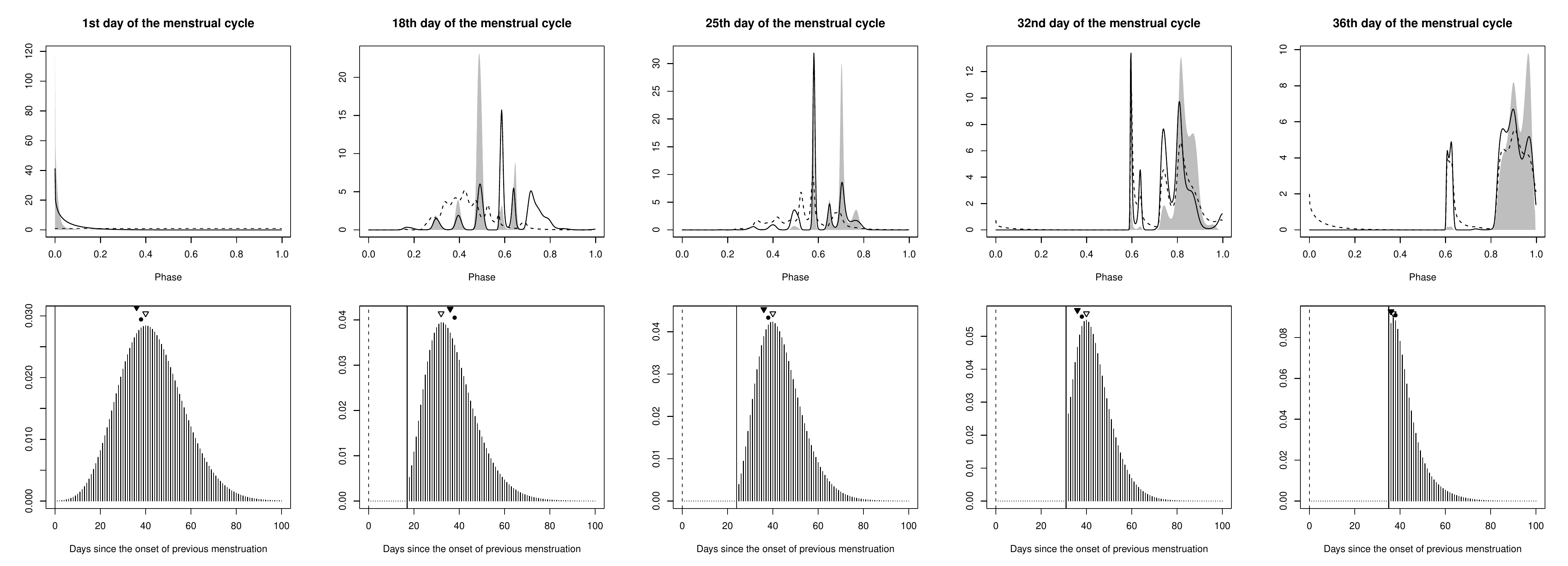}
\end{center}
\caption{
  Example estimated conditional distributions for menstrual phase (upper panels)
  and the associated predictive distributions for the day of onset of menstruation (lower panels).
  In the upper panels, predictive and filtering distributions
  are shown as dashed and solid lines, respectively, and smoothed distributions
  are represented by gray shading.
  $x$-axes represent $\theta_t \mod 1$.
  In the lower panels, the marginal probabilities of the onset of menstruation $h(k \mid Y_t, Z_t)$
  are shown.
  Dashed and solid vertical lines indicate the day of onset of the previous menstruation
  and the current day, respectively.
  Filled circles indicate the actual day of onset of the next menstruation.
  Filled and open triangles indicate the best conventional prediction for the subject
  and the model-based prediction, respectively.
  These results were obtained by applying the best AIC model to 
  the first menstrual cycle of the test data of subject 1.
  From left to right, panels correspond to 21, 14, 7, and 3 days before
  the upcoming day of onset of menstruation, respectively.
}
\label{fig:phase}
\end{sidewaysfigure*}

Parameter estimates in the best AIC model for the 10 subjects are shown
in Table \ref{table:estimates}.
The selected order of the trigonometric series ranged from 5 to 12.
The estimated relationship between expected BBT and menstrual phase, 
and the estimated probability density distribution for phase advancement
for each subject are shown in Figure \ref{fig:all}.
Although the estimated temperature-phase regression lines were ``squiggly''
due to the high order of the trigonometric series,
they in general exhibited two distinct stages:
the temperature tended to be lower in the first half of the cycle and
higher in the second half, which is consistent with the well-known
periodicity of BBT through the menstrual cycle \citep{Barron2005}.
Figure \ref{fig:phase} shows examples of the estimated conditional distributions
for menstrual phase and the associated predictive distributions for
the day of onset of menstruation.

The RMSE of the predictions of the next day of onset of menstruation 
provided by the sequential method and by the conventional method
are shown in Figure \ref{fig:rmse} (see Tables 
\ref{table:rmse_seq} and \ref{table:rmse_stat} in Appendix \ref{sect:appb} for more details).
In the sequential method, the RMSE tended to decrease
as the day approached the next day of onset of menstruation, 
suggesting that accumulation of the BBT time series data
contributed to increasing the accuracy of the prediction.
Although, at the day of onset of preceding menstruation,
the sequential method does not necessarily provide a more accurate prediction
compared with the conventional method,
it may provide a much-improved prediction as the day gets closer to
the onset of next menstruation.
Except for in subject 9, the sequential method exhibited a better predictive performance 
than the conventional method for at least one of the time points of prediction we considered
(i.e., the day preceding the next day of onset of menstruation) (Figure \ref{fig:rmse}).
Compared with the best prediction provided by the conventional method, 
the range, mean, and median of the rate of the maximum reduction in the RMSE 
for each subject in the sequential method was 0.066--1.481, 0.557 and 0.487, respectively.
Note that the predictive performance tended to be even better
when the accuracy was measured by using the mean absolute error (see 
Tables \ref{table:mae_seq} and \ref{table:mae_stat} and Figure \ref{fig:mae} 
in Appendix \ref{sect:appb}).

\begin{sidewaystable*}[p]
  \centering
  \caption{Results of fitting the state-space model by using the maximum likelihood method.
    Parameter estimates and their 95\% confidence intervals (in brackets) of the best AIC model are shown for each subject.
    Log-likelihood was approximated by using the non-Gaussian filter with the state-space was discretized with 512 intervals.
    }
  {\tiny
  \begin{tabular}{lcccccccccc} \toprule
    & \multicolumn{10}{c}{Subject} \\
     & 1 & 2 & 3 & 4 & 5 & 6 & 7 & 8 & 9 & 10 \\
    \midrule
    Best model       & M10 & M8 & M9 & M6 & M11 & M8 & M11 & M12 & M10 & M5  \\
    $\alpha$  & 0.210 & 0.953 & 0.344 & 1.520 & 0.318 & 1.971 & 0.630 & 0.172 & 0.150 & 0.201 \\
     & $[0.151, 0.292]$ & $[0.628,  1.439]$ & $[0.253, 0.467]$ & $[1.015,  2.266]$ & $[0.233, 0.435]$ & $[1.329,  2.905]$ & $[0.449, 0.882]$ & $[0.128, 0.231]$ & $[0.106, 0.211]$ & $[0.140, 0.288]$ \\
    $\beta$  & 8.915 & 32.131 & 10.916 & 45.146 & 8.536 & 59.929 & 19.510 & 5.886 & 5.284 & 6.544 \\
     & $[6.303, 12.610]$ & $[21.157, 48.798]$ & $[7.883, 15.115]$ & $[30.420, 67.002]$ & $[6.094, 11.955]$ & $[40.536, 88.601]$ & $[13.806, 27.572]$ & $[4.236, 8.178]$ & $[3.595, 7.766]$ & $[4.429, 9.669]$ \\
    $\sigma$   & 0.112 & 0.161 & 0.119 & 0.121 & 0.093 & 0.152 & 0.108 & 0.101 & 0.116 & 0.209 \\
     & $[0.103, 0.121]$ & $[0.150,  0.173]$ & $[0.109, 0.130]$ & $[0.113,  0.129]$ & $[0.085, 0.102]$ & $[0.142,  0.163]$ & $[0.098, 0.120]$ & $[0.094, 0.109]$ & $[0.109, 0.123]$ & $[0.195, 0.224]$ \\
    $a$   & 36.299 & 36.203 & 36.485 & 36.384 & 36.632 & 36.522 & 36.518 & 36.519 & 36.645 & 36.123 \\
     & $[36.247, 36.351]$ & $[36.187, 36.219]$ & $[36.466, 36.504]$ & $[36.370, 36.397]$ & $[36.614, 36.650]$ & $[36.509, 36.535]$ & $[36.501, 36.536]$ & $[36.492, 36.546]$ & $[36.620, 36.670]$ & $[36.087, 36.160]$ \\
    $b_1$   & 0.193 & 0.197 & 0.174 & 0.098 & $-0.034$ & 0.107 & 0.130 & 0.141 & 0.129 & 0.138 \\
     & $[0.153, 0.232]$ & $[0.179,  0.216]$ & $[0.147, 0.200]$ & $[0.074,  0.122]$ & $[-0.072, 0.003]$ & $[0.088,  0.126]$ & $[0.101, 0.159]$ & $[0.114, 0.169]$ & $[0.099, 0.159]$ & $[0.082, 0.193]$ \\
    $b_2$   & 0.015 & 0.024 & $-0.028$ & $-0.079$ & 0.115 & $-0.063$ & $-0.001$ & $-0.370$ & $-0.012$ & 0.002 \\
     & $[-0.021, 0.050]$ & $[-0.010, 0.058]$ & $[-0.061, 0.006]$ & $[-0.096, -0.061]$ & $[0.063, 0.167]$ & $[-0.081, -0.046]$ & $[-0.024, 0.022]$ & $[-0.416, -0.323]$ & $[-0.067, 0.043]$ & $[-0.082, 0.085]$ \\
    $b_3$   & 0.009 & $-0.057$ & 0.027 & 0.032 & $-0.098$ & 0.018 & 0.025 & $-0.089$ & $-0.087$ & $-0.126$ \\
     & $[-0.112, 0.130]$ & $[-0.085, -0.029]$ & $[-0.001, 0.056]$ & $[0.012,  0.051]$ & $[-0.148, -0.048]$ & $[-0.003, 0.039]$ & $[0.001, 0.049]$ & $[-0.132, -0.045]$ & $[-0.136, -0.038]$ & $[-0.192, -0.060]$ \\
    $b_4$   & $-0.039$ & 0.021 & $-0.023$ & 0.007 & 0.124 & $-0.028$ & $-0.032$ & 0.167 & $-0.078$ & $-0.153$ \\
     & $[-0.075, -0.003]$ & $[-0.010, 0.053]$ & $[-0.051, 0.005]$ & $[-0.017, 0.030]$ & $[0.094, 0.153]$ & $[-0.054, -0.001]$ & $[-0.061, -0.003]$ & $[0.105, 0.229]$ & $[-0.125, -0.030]$ & $[-0.250, -0.056]$ \\
    $b_5$   & 0.058 & $-0.046$ & 0.013 & $-0.020$ & $-0.129$ & $-0.030$ & 0.029 & 0.174 & 0.049 & $-0.003$ \\
     & $[-0.006, 0.121]$ & $[-0.097, 0.005]$ & $[-0.028, 0.054]$ & $[-0.039, -0.002]$ & $[-0.165, -0.092]$ & $[-0.061, 0.000]$ & $[0.001, 0.057]$ & $[0.137, 0.212]$ & $[-0.040, 0.139]$ & $[-0.153, 0.148]$ \\
    $b_6$   & $-0.014$ & 0.046 & $-0.065$ & $-0.022$ & 0.097 & 0.028 & $-0.001$ & $-0.038$ & 0.100 &  NA \\
     & $[-0.230, 0.202]$ & $[-0.005, 0.098]$ & $[-0.106, -0.024]$ & $[-0.051, 0.006]$ & $[0.045, 0.149]$ & $[-0.001, 0.058]$ & $[-0.030, 0.029]$ & $[-0.122, 0.047]$ & $[0.040, 0.160]$ &    \\
    $b_7$   & $-0.024$ & 0.019 & 0.006 &   NA & $-0.078$ & $-0.026$ & 0.024 & $-0.112$ & $-0.003$ &      NA \\
     & $[-0.485, 0.437]$ & $[-0.030, 0.068]$ & $[-0.034, 0.046]$ &   & $[-0.124, -0.032]$ & $[-0.068, 0.016]$ & $[-0.015, 0.062]$ & $[-0.149, -0.075]$ & $[-0.104, 0.099]$ &   \\
    $b_8$   & 0.037 & 0.034 & $-0.063$ &       NA & 0.088 & 0.037 & 0.060 & $-0.098$ & $-0.058$ &      NA \\
     & $[-0.256, 0.331]$ & $[-0.037, 0.105]$ & $[-0.145, 0.020]$ &   & $[0.047, 0.129]$ & $[-0.011, 0.085]$ & $[0.011, 0.109]$ & $[-0.180, -0.017]$ & $[-0.171, 0.056]$ &  \\
    $b_9$   & $-0.086$ &   NA & $-0.045$ &      NA & $-0.060$ &   NA & 0.007 & 0.010  & $-0.036$ &      NA  \\
     & $[-0.231, 0.060]$ &  &$[-0.142, 0.051]$ &  &$[-0.090, -0.029]$ &   &$[-0.019, 0.033]$ & $[-0.033, 0.053]$ & $[-0.136, 0.064]$ &   \\
    $b_{10}$   & 0.061 &       NA &   NA &       NA & 0.101 &       NA & 0.050 & 0.120 & 0.053 &      NA \\
     & $[-0.170, 0.293]$ &  &   &  &$[0.062, 0.140]$ &  & $[0.002, 0.097]$ & $[0.062, 0.178]$ & $[-0.036, 0.141]$ &  \\
    $b_{11}$   & NA &       NA &       NA &       NA & $-0.058$ &       NA & 0.031 & 0.023 &    NA &     NA  \\
     &  &  &  &  & $[-0.100, -0.016]$ &  & $[-0.032, 0.094]$ & $[-0.023, 0.069]$ &   &  \\
    $b_{12}$   & NA &       NA &       NA &       NA &   NA &       NA &    NA & $-0.114$ &       NA &      NA \\
     & &  &  &  &  &  &   & $[-0.145, -0.082]$ &   &   \\
    $c_1$   & $-0.193$ & $-0.108$ & $-0.182$ & $-0.270$ & $-0.184$ & $-0.170$ & $-0.254$ & $-0.417$ & $-0.172$ & $-0.276$ \\
     & $[-0.304, -0.081]$ & $[-0.140, -0.075]$ & $[-0.218, -0.146]$ & $[-0.289, -0.252]$ & $[-0.204, -0.164]$ & $[-0.190, -0.150]$ & $[-0.274, -0.233]$ & $[-0.463, -0.371]$ & $[-0.218, -0.126]$ & $[-0.334, -0.217]$ \\
    $c_2$   & $-0.051$ & $-0.131$ & $-0.130$ & $-0.034$ & $-0.107$ & $-0.058$ & $-0.038$ & $-0.216$ & $-0.035$ & $-0.057$  \\
     & $[-0.085, -0.018]$ & $[-0.154, -0.108]$ & $[-0.159, -0.101]$ & $[-0.058, -0.010]$ & $[-0.133, -0.080]$ & $[-0.082, -0.035]$ & $[-0.059, -0.017]$ & $[-0.263, -0.169]$ & $[-0.069, -0.001]$ & $[-0.107, -0.006]$ \\
    $c_3$   & $-0.089$ & $-0.029$ & $-0.022$ & $-0.056$ & 0.006 & $-0.023$ & $-0.021$ & 0.062 & $-0.070$ & $-0.019$  \\
     & $[-0.120, -0.059]$ & $[-0.065, 0.008]$ & $[-0.051, 0.006]$ & $[-0.076, -0.037]$ & $[-0.041, 0.052]$ & $[-0.046, -0.000]$ & $[-0.044, 0.002]$ & $[0.018, 0.106]$ & $[-0.126, -0.014]$ & $[-0.096, 0.059]$  \\
    $c_4$   & $-0.030$ & $-0.001$ & $-0.035$ & 0.005 & $-0.016$ & $-0.012$ & 0.016 & 0.248 & $-0.047$ & 0.042 \\
     & $[-0.174, 0.114]$ & $[-0.042, 0.041]$ & $[-0.063, -0.007]$ & $[-0.014, 0.025]$ & $[-0.077, 0.046]$ & $[-0.038, 0.014]$ & $[-0.013, 0.044]$ & $[0.202, 0.295]$ & $[-0.118, 0.024]$ & $[-0.065, 0.148]$ \\
    $c_5$   & $-0.003$ & $-0.058$ & 0.027 & $-0.017$ & 0.010 & $-0.037$ & $-0.016$ & $-0.040$ & $-0.082$ & $-0.126$ \\
     & $[-0.256, 0.250]$ & $[-0.117, 0.001]$ & $[-0.007, 0.062]$ & $[-0.040, 0.007]$ & $[-0.047, 0.067]$ & $[-0.067, -0.007]$ & $[-0.044, 0.011]$ & $[-0.099, 0.019]$ & $[-0.134, -0.029]$ & $[-0.186, -0.066]$ \\
    $c_6$   & $-0.053$ & $-0.003$ & 0.023 & $-0.034$ & $-0.007$ & $-0.025$ & $-0.010$ & $-0.277$ & 0.004 & NA \\
     & $[-0.143, 0.037]$ & $[-0.055, 0.050]$ & $[-0.030, 0.076]$ & $[-0.059, -0.008]$ & $[-0.052, 0.038]$ & $[-0.062, 0.013]$ & $[-0.036, 0.015]$ & $[-0.312, -0.243]$ & $[-0.107, 0.114]$ &   \\
    $c_7$   & 0.068 & $-0.030$ & $-0.031$ &   NA & 0.010 & $-0.040$ & $-0.052$ & $-0.075$ & 0.059 &      NA  \\
     & $[-0.040, 0.176]$ & $[-0.080, 0.020]$ & $[-0.073, 0.012]$ &  & $[-0.038, 0.057]$ & $[-0.086, 0.006]$ & $[-0.081, -0.023]$ & $[-0.136, -0.015]$ & $[0.014,  0.104]$ &  \\
    $c_8$   & $-0.050$ & 0.044 & 0.067 &       NA & $-0.044$ & 0.027 & 0.027 & 0.214 & 0.074 &      NA  \\
     & $[-0.324, 0.223]$ & $[-0.002, 0.089]$ & $[0.006,  0.127]$ &  & $[-0.102, 0.013]$ & $[-0.028, 0.082]$ & $[-0.014, 0.068]$ & $[0.171, 0.257]$ & $[-0.009, 0.157]$ &   \\
    $c_9$   & $-0.027$ &   NA & $-0.085$ &      NA & 0.006 &   NA & 0.007 & 0.079 & $-0.067$ &      NA  \\
     & $[-0.675, 0.620]$ &  & $[-0.134, -0.036]$ &  & $[-0.059, 0.071]$ &   & $[-0.049, 0.063]$ & $[0.036, 0.122]$ & $[-0.169, 0.034]$ &   \\
    $c_{10}$   & 0.025 &       NA &   NA &       NA & $-0.019$ &       NA & 0.013 & $-0.107$ & $-0.057$ &      NA  \\
     & $[-0.494, 0.543]$ &  &  &  & $[-0.088, 0.050]$ &  & $[-0.032, 0.058]$ & $[-0.172, -0.041]$ & $[-0.156, 0.042]$ &  \\
    $c_{11}$   & NA &       NA &       NA &       NA & 0.041 &       NA & 0.067 & $-0.022$ &   NA &      NA \\
     &  &  &  &  & $[-0.011, 0.093]$ &  & $[0.025,  0.108]$ & $[-0.057, 0.014]$ &   &   \\
    $c_{12}$   & NA &       NA &       NA &       NA &    NA &       NA &    NA & 0.012 &       NA &      NA \\
     &  &  &  &  &  &  &  & $[-0.053, 0.077]$ &  &  \\
    log-likelihood   & $468.905$ & $164.278$ & $206.631$ & $415.909$ & $343.887$ & $234.46$ & $354.47$ & $408.464$ & $330.137$ & $-178.022$  \\
    \bottomrule
  \end{tabular}
  }
  \label{table:estimates}
\end{sidewaystable*}

\section{Discussion}
\label{sect:disc}

Here we constructed a statistical framework 
that provides a model-based prediction of the day of onset of menstruation 
based on a state-space model and an associated Bayesian filtering algorithm.
The model describes the daily fluctuation of BBT and the history of menstruation.
The filtering algorithms yielded a filtering distribution of menstrual phase 
that was conditional on all of the data available at that point in time, 
which was used to derive a predictive distribution for the next day of onset of menstruation.

The predictive framework we developed has several
notable characteristics that make it superior to the conventional method
of calendar calculation for predicting the day of onset of menstruation.
State-space modeling and Bayesian filtering techniques
enable the proposed method to yield sequential predictions
of the next day of onset of menstruation based on daily BBT data.
Even though menstrual cycle length fluctuates stochastically,
the prediction of day of onset of menstruation is automatically adjusted 
based on the daily updated filtering distribution, yielding a
flexible yet robust prediction of the next day of onset of menstruation.
This is markedly different from the conventional method that
yields only a fixed prediction that is never adjusted.
In a study similar to the present study, \cite{Bortot2010} constructed a predictive framework
for menstrual cycle length by using the state-space modeling approach.
However, in their model, 
the one-step-ahead predictive distribution for menstrual cycle length
is obtained based on the past time series of cycle length
and within-cycle information is not taken into account in the prediction.

\begin{figure*}[p]
{\centering
\includegraphics[bb=0 0 432 288,height=2in]{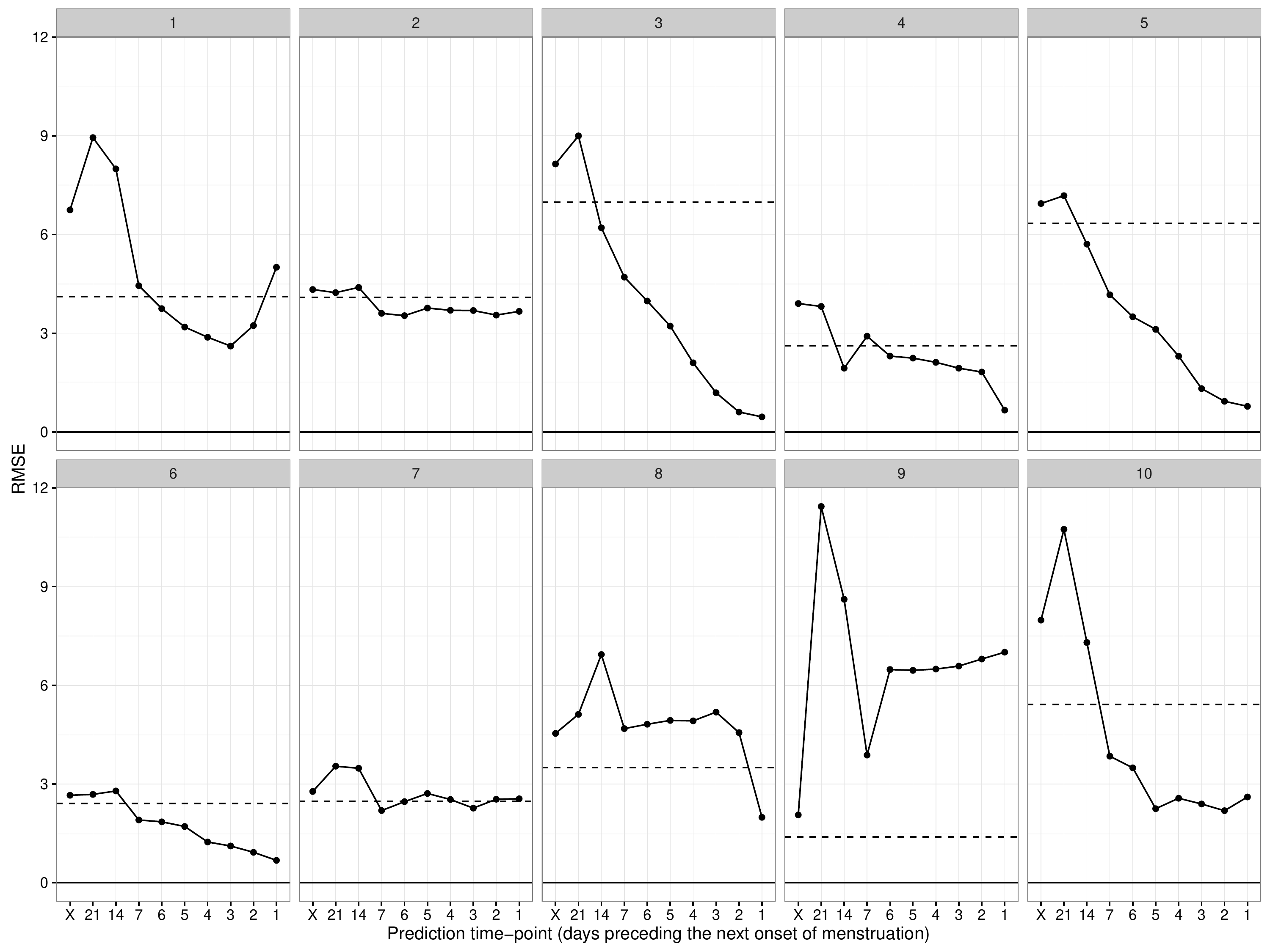}
}
\caption{
  Root mean square error (RMSE) of the prediction of the day of
  onset of the next menstruation.
  Solid lines indicate the RMSE of the sequential prediction
  (based on the best AIC model for each subject),
  and horizontal dashed lines indicate the lowest RMSE of the conventional
  prediction for each subject.
  On the horizontal axis, ``X'' indicates the prediction obtained at the day
  of onset of the previous menstruation.
  Each panel shows the results for one subject.
}
\label{fig:rmse}
\end{figure*}

Furthermore, compared to the conventional method, the proposed sequential method
generally yielded a more accurate prediction of day of onset of menstruation.
As the day approaches the next onset of menstruation and more daily temperature data
are accumulated, the proposed method produced a considerably 
improved prediction (Figure \ref{fig:rmse}).
Since measuring BBT is a simple and inexpensive means of determining the
current phase of the menstrual cycle, the proposed framework can provide predictions 
of the next onset of menstruation that can easily be implemented for 
the management of women's health.
Although the non-Gaussian filtering technique may be 
computationally impractical for a state-space model with a
high-dimensional state vector \citep{Kitagawa1987}, 
our proposed model only involves a two-dimensional state vector
so the computational cost required for filtering single data
is negligible for a contemporary computer.

The conditional probability density distribution for menstrual phase obtained with
the sequential Bayesian filter may also be used for
predicting other events that are related to the menstrual cycle.
For example, the conditional probability of
the menstrual phase being below or above a ``change point'' of expected temperature
could be used as a model-based probability of the subject being 
in the follicular phase (before ovulation) or the luteal phase (after ovulation).
These probabilities may also be used to estimate the timing of ovulation.
Combined with a fecundability model that provides the probability of
conception within a menstrual cycle that is conditional on daily intercourse behavior,
the model might also be used to predict the probability of conception \citep{Bortot2010, Lum2015}.
As the Bayesian filtering algorithm can yield predictive and smoothed distributions
(Equations \ref{eqn:predictor} and \ref{eqn:smoother}),
these predictions could be obtained in both a prospective and a retrospective manner.
Another possible extension of the model may be the addition of a new observation model
for the physical condition of subjects, which may help to estimate the menstrual phase more
precisely or to provide a forecast of the physical condition of subject as their 
menstrual cycle progresses.

We note that the state-space model proposed in the present study may not be 
sufficiently flexible to describe the full variation in menstrual cycle length.
In the present study, we fitted our model only to data from 10 subjects 
and this data did not include extremely short or extremely long cycles
because a preliminary investigation suggested that 
inclusion of such data could result in unreasonable parameter estimates (results not shown).
Specifically, the estimates of $\alpha$ and $\beta$ may become extremely small,
leading to an almost flat probability density distribution for menstrual phase advancement.
Under these parameter values, the predictive distribution for the onset of menstruation
also becomes flat, preventing a useful prediction from being obtained.
These results suggest that with a single set of parameter values, 
the proposed state-space model does not capture the whole observed variation
in cycle length.
It is known that the statistical distribution of menstrual cycle length
is characterized by a mixture distribution that comprises standard 
and nonstandard cycles, where, for the latter, a skewed distribution may well
represent the observed pattern \citep[e.g.,][]{Harlow1991,Guo2006, Lum2015}.
Our proposed model, however, does not provide such a mixture-like marginal
distribution for the day of onset of menstruation.
It is also known that the mean and variance of the marginal distribution of cycle length
can vary depending on subjects' age \citep{Guo2006, Bortot2010, Huang2014}.
Therefore, we assume that modeling variations in the system model parameters
(i.e., $\alpha$ and $\beta$),
by including covariates such as age or
within- and among-subject random effects, or both,
may be a promising extension of the proposed framework.
To explain the skewed marginal distribution of menstrual cycle length,
\cite{Sharpe1980} considered a rate process in which the development rate fluctuates randomly.
Inclusion of random effects for the system model parameters would allow us
to accommodate this idea.
Although the inclusion of random effects will increase the flexibility of the framework,
it would also considerably complicate the likelihood calculation and make
parameter estimation more challenging.

The state-space model proposed here provides a conceptual description of the menstrual cycle.
We explain this perspective by using the analogy of a clock that
makes one complete revolution in each menstrual cycle.
The system model expresses 
the hand of the clock (i.e., the latent phase variable) moving steadily
forward with an almost steadily, yet slightly variable, pace.
Observation models describe observable events that are imprinted on the clock's dial;
for example, menstruation is scheduled to occur when the hand of this clock arrives
at a specific point on the dial.
The fluctuation in BBT, as well as other possibly related phenomena such as ovulation,
would also be marked on the dial.
Even though the position of the hand of the clock is unobservable, 
we can estimate it as a conditional distribution of the latent phase variable 
by using the Bayesian filtering technique,
which enables us to make a sequential, model-based prediction of menstruation.
Although this is a rather phenomenological view of the menstrual cycle,
it is useful for developing a rigorous and extendable modeling framework for 
predicting and studying phenomena that are associated with the menstrual cycle.

%

\section*{Acknowledgements}

We are grateful to K. Shimizu, T. Matsui, A. Tamamori,
M. L. Taper and J. M. Ponciano
who provided valuable comments on this research.
These research results were achieved by ``Research and Development on Fundamental
and Utilization Technologies for Social Big Data'', the Commissioned Research of National
Institute of Information and Communications Technology (NICT), Japan.
This research was initiated by an application to the ISM
Research Collaboration Start-up program (No RCSU2013--08) from M. Kitazawa.





\clearpage
\newpage

\appendix

\section{State estimation, likelihood calculation, and prediction using the non-Gaussian filter}
\label{sect:appa}

The statistical inferences described in Section \ref{sect:model} of the main text 
were based on an estimation of the conditional distribution
of the latent menstrual phase.
Although the conditional distribution can be obtained by using 
the recursive equations described in Section \ref{sect:ngf},
the integrals in those formulae are generally not analytically tractable
for non-linear, non-Gaussian, state-space models.
Therefore, to obtain approximate conditional distributions,
filtering algorithms for general state-space models are required.
We use the non-Gaussian filter proposed by \cite{Kitagawa1987},
in which the continuous state space is discretized with 
equally spaced fixed points at which the conditional probability density is evaluated.
With these discretized probability density functions,
integrals are evaluated by using numerical integration.
This filtering algorithm also yields an approximate log-likelihood,
making the maximum likelihood estimation of the model parameters possible.
Detailed descriptions for this numerical procedure are provided, for example,
in \cite{Kitagawa1987, Kitagawa2010} and \cite{deValpine2002}.

To apply non-Gaussian filtering to the state-space model described in Section \ref{sect:ssm},
it is computationally more convenient to consider the state space of a
circular phase variable rather than a linear phase variable described in the main text,
because the latter requires discretization of unbounded real space.
Hence, in the following, we first reformulate the statistical modeling framework
in terms of a circular latent phase variable.
The reformulated model conceptually agrees with the original formulation, 
given the probability that the menstrual phase advancement per day being more than 1
(i.e., $\mbox{Pr}(\epsilon_t > 1)$) is negligible.
In practice, this condition is naturally attained in the 
estimation of the model parameters
as long as unusually short menstrual cycles (e.g., 3 to 4 days)
are not included in the dataset.
Below, we provide a numerical procedure to obtain the conditional distributions,
log-likelihood, and the predictive distribution of the next day of onset of menstruation
based on the non-Gaussian filtering technique (note, all definitions and notations are
the same as those used in Section \ref{sect:model} of the main text).

\subsection{Model with a circular state variable}

Instead of using the real latent state variable for the menstrual phase,
$\theta_t \in \mathbb{R}$, we reformulate the model with 
a circular latent variable of period 1, denoted by $\omega_t\in [0, 1)$.
We consider that the correspondence between these two variables is 
characterized by the following projection:
$\omega_t = f(\theta_t) = \theta_t \mod 1$.
The system model for this circular variable is then expressed as
%
\begin{align}
  & \omega_t = \left( \omega_{t-1} + \epsilon_t \right) \mod 1 \\
  & \epsilon_t \sim \textrm{Gamma}(\alpha, \beta).
\end{align}
Under this assumption, given $\omega_{t-1}$, the conditional distribution of $\omega_t$ 
is a wrapped gamma distribution with a probability density function:
%
\begin{align}
  &p(\omega_t \mid \omega_{t-1}) \notag \\
  &= \textrm{wGamma}(\alpha, \beta) \notag \\
  & = \frac{\beta^\alpha}{\Gamma(\alpha)}
  \exp \left\{-\beta \delta(\omega_t, \omega_{t-1}) \right\} \notag \\
  & \hspace{4em} \times \Phi\left\{\exp(-\beta), 1-\alpha, \delta(\omega_t, \omega_{t-1})\right\},
  \label{eqn:system}
\end{align}

\noindent{where} $\delta(\omega_t, \omega_{t-1}) = (\omega_t - \omega_{t-1})^{I(\omega_t > \omega_{t-1})}(1 + \omega_t - \omega_{t-1})^{I(\omega_t \leq \omega_{t-1})}$ is the advancement of the menstrual phase
on the circular scale and
$\Phi(z, s, a)=\sum_{k=0}^{\infty}\frac{z^k}{(a+k)^s}$
is Lerch's transcendental function \citep{Coelho2007}.

The observation model for BBT, which is conditional on $\omega_t$, is expressed as
%
\begin{align}
  y_t &= a + \sum_{m=1}^M 
      (b_m\cos2m\pi\omega_t + c_m\sin2m\pi\omega_t) + e_t \hspace{2em} \\
  e_t &\sim \textrm{Normal}(0, \sigma^2),
\end{align}
\noindent{leading} to a normal conditional distribution of $y_t$ with a
probability density function:
%
\begin{align}
  p(y_t \mid \omega_t) &= \textrm{Normal}\left\{\mu(\omega_t), \sigma^2 \right\} \notag \\
  & = \frac{1}{\sqrt{2\pi\sigma^2}}
  \exp \left[ -\frac{\left\{y_t - \mu(\omega_t)\right\}^2}{2\sigma^2} \right],
\end{align}
\noindent{where} $\mu(\omega_t) = a + \sum_{m=1}^M 
      (b_m\cos2m\pi\omega_t + c_m\sin2m\pi\omega_t)$.

Assuming that $\mbox{Pr}(\epsilon_t > 1)$ is negligible, 
we consider menstruation to have started when $\omega_t$ ``steps over'' $1$.
This is represented as
%
\begin{align}
  z_t &= 0 \hspace{2em} \textrm{when} \hspace{2em} \omega_t > \omega_{t-1} \\
  &= 1 \hspace{2em} \textrm{when} \hspace{2em} \omega_t \leq \omega_{t-1}. 
\end{align}
We can write this deterministic allocation in a probabilistic manner
that is conditional on $(\omega_t, \omega_{t-1})$,
where $z_t$ follows a Bernoulli distribution:
%
\begin{align}
  &p(z_t \mid \omega_t, \omega_{t-1}) = \notag \\
  &(1 - z_t)\left\{I(\omega_t > \omega_{t-1})\right\}+ z_t\left\{I(\omega_t \leq \omega_{t-1})\right\}.
\end{align}

The log-likelihood of this model can then be expressed as
%
\begin{align}
  &l(\boldsymbol{\xi}; Y_T, Z_T)= \notag \\
  &\log p(y_1,z_1 \mid \boldsymbol{\xi})+
  \sum_{t=2}^T\log p(y_t,z_t \mid Y_{t-1}, Z_{t-1}, \boldsymbol{\xi}),
\end{align}
\noindent{where}
%
\begin{align}
  &\log p(y_1,z_1 \mid \boldsymbol{\xi})  \notag \\
  &= \log \int_0^1\int_0^1 p(y_1 \mid \omega_1)p(z_1 \mid \omega_1, \omega_0)p(\omega_1, \omega_0) d\omega_1 d\omega_0,  
\end{align}
\noindent{and} for $t = 2, \dots, T$,
%
\begin{align}
  &\log p(y_t,z_t \mid Y_{t-1}, Z_{t-1}, \boldsymbol{\xi}) \notag \\
  &= \log \int_0^1\int_0^1 p(y_t \mid \omega_t)p(z_t \mid \omega_t, \omega_{t-1}) \notag \\
  & \hspace{4em}\times p(\omega_t, \omega_{t-1} \mid Y_{t-1}, Z_{t-1}) d\omega_t d\omega_{t-1}.  
\end{align}

For $t=1,\dots,T$, the prediction, filtering, and smoothing equations for the above model are expressed as follows:

\vspace{-.2em}
{\scriptsize
\begin{align}
  &p(\omega_t, \omega_{t-1} \mid Y_{t-1}, Z_{t-1}) \notag \\
  & = p(\omega_t \mid \omega_{t-1})p(\omega_{t-1} \mid Y_{t-1},Z_{t-1}) \notag \\ 
  & = p(\omega_t \mid \omega_{t-1}) \int_0^1 p(\omega_{t-1}, \omega_{t-2} \mid Y_{t-1},Z_{t-1}) d\omega_{t-2} \\
  &p(\omega_t, \omega_{t-1} \mid Y_t, Z_t) \notag \\
  & = \frac{p(y_t, z_t \mid \omega_t, \omega_{t-1})p(\omega_t, \omega_{t-1} \mid Y_{t-1}, Z_{t-1})}{\int_0^1\int_0^1 p(y_t, z_t \mid \omega_t, \omega_{t-1})p(\omega_t, \omega_{t-1} \mid Y_{t-1}, Z_{t-1}) d\omega_t d\omega_{t-1}}\notag \\
  & = \frac{p(y_t \mid \omega_t)p(z_t \mid \omega_t, \omega_{t-1})p(\omega_t, \omega_{t-1} \mid Y_{t-1}, Z_{t-1})}{\int_0^1\int_0^1 p(y_t \mid \omega_t)p(z_t \mid \omega_t, \omega_{t-1})p(\omega_t, \omega_{t-1} \mid Y_{t-1}, Z_{t-1}) d\omega_t d\omega_{t-1}}  \\
  & p(\omega_t, \omega_{t-1} \mid Y_T, Z_T) \notag \\
    & = p(\omega_t, \omega_{t-1} \mid Y_t, Z_t)
      \int_0^1 \frac{p(\omega_{t+1}, \omega_t \mid Y_T, Z_T)
        p(\omega_{t+1} \mid \omega_t)}
        {p(\omega_{t+1}, \omega_t \mid Y_t, Z_t)} d\omega_{t+1} \notag \\
    & = \frac{p(\omega_t, \omega_{t-1} \mid Y_t, Z_t) \int_0^1 p(\omega_{t+1}, \omega_t \mid Y_T, Z_T) d\omega_{t+1}}
  {p(\omega_t \mid Y_t, Z_t)} \notag \\
    & = \frac{p(\omega_t, \omega_{t-1} \mid Y_t, Z_t) \int_0^1 p(\omega_{t+1}, \omega_t \mid Y_T, Z_T) d\omega_{t+1}}
    {\int_0^1 p(\omega_t, \omega_{t-1} \mid Y_t, Z_t) d\omega_{t-1}} ,
\end{align}
}
\noindent{where} for $t = 1$, we set $p(\omega_1, \omega_0 \mid Y_0, Z_0)$
as $p(\omega_1, \omega_0)$, which is the specified initial distribution for the phase.

We consider the conditional probability that 
the next menstruation has started by day $k + t$,
given the phase state $\omega_t$, to be
$F(k \mid \omega_t) = \textrm{Pr}\left\{\Delta_k(t) > 1 - \omega_t\right\}$.
Then, $F(k \mid \omega_t)$ represents the conditional distribution function 
for the onset of menstruation, and under the assumption of the 
state-space model described above, this is given as
%
\begin{align}
  F(k \mid \omega_t) &= \int_{1-\omega_t}^{\infty} 
    g(x; k\alpha, \beta) dx \notag \\
  &= 1 - G(1-\omega_t; k\alpha, \beta).
\end{align}
The conditional probability function for the day of onset of menstruation,
$f(k \mid \omega_t)$, is then given as
%
\begin{align}
  f(k \mid\omega_t) &= F(k \mid \omega_t) - F(k-1 \mid \omega_t) \notag \\
  &= \left\{ 1 - G(1-\omega_t; k\alpha, \beta) \right\} \notag \\
  &\hspace{4em}- \left[ 1 - G\left\{1-\omega_t; (k-1)\alpha, \beta\right\} \right] \notag \\
  &=  G\left\{1-\omega_t; (k-1)\alpha, \beta\right\} 
      - G(1-\omega_t; k\alpha, \beta),
\end{align}
\noindent{where} we set $F(0 \mid \omega_t)=0$.

The marginal distribution for the day of onset of menstruation, $h(k \mid Y_t, Z_t)$,
is then obtained with the marginal filtering distribution for the phase state,
$p(\omega_t \mid Y_t, Z_t)$, which is given as
%
\begin{equation}
  h(k \mid Y_t, Z_t) = \int_0^1 f(k \mid \omega_t)p(\omega_t \mid Y_t, Z_t) d\omega_t,
\end{equation}
\noindent{which} is used to provide a point prediction for the day of onset of menstruation
by finding the value of $k$ that gives the highest probability, $\max h(k \mid Y_t, Z_t)$.

\subsection{Numerical procedure for non-Gaussian filtering}

We let $\omega(i), ~i=1,\dots, N$ be equally spaced grid points with interval $[0, 1)$.
We denote the approximate probability density of the
predictive, filtering, and smoothed distributions evaluated at these
grid points on the state space by
$\tilde{p}_{\mbox{p}}(i, j, t), \tilde{p}_{\mbox{f}}(i, j, t)$, and
$\tilde{p}_{\mbox{s}}(i, j, t)$, which are defined respectively as
%
\begin{align}
  &\tilde{p}_{\mbox{p}}(i, j, t) 
    = p\left\{\omega_t=\omega(i), \omega_{t-1}=\omega(j) \mid Y_{t-1}, Z_{t-1}\right\}  \\
  &\tilde{p}_{\mbox{f}}(i, j, t) 
    = p\left\{\omega_t=\omega(i), \omega_{t-1}=\omega(j) \mid Y_t, Z_t\right\}  \\
  &\tilde{p}_{\mbox{s}}(i, j, t) 
    = p\left\{\omega_t=\omega(i), \omega_{t-1}=\omega(j) \mid Y_T, Z_T\right\},
\end{align}
\noindent{where} for $i,j = 1,\dots,N$ and $t=1,\dots,T$.
As the initial distribution, we also specify the probability density 
$\tilde{p}_{\mbox{p}}(i, j, 0)$.
We define the probability density function of the system model (Equation \ref{eqn:system})
evaluated at each grid point as
$\tilde{p}_{\mbox{m}}(i, j) = p\left\{\omega_t = \omega(i) \mid \omega_{t-1} = \omega(j)\right\}$.
Then, the prediction, filtering, and smoothing equations, respectively, are 
expressed as 
%
\begin{equation}
  \tilde{p}_{\mbox{p}}(i, j, t) 
  = \tilde{p}_{\mbox{m}}(i, j)\tilde{p}_{\mbox{f}}'(j, t-1)
\end{equation}
\begin{align}
  &\tilde{p}_{\mbox{f}}(i, j, t) \notag \\
  &= \frac{p\left\{y_t \mid \omega_t=\omega(i)\right\}p\left\{z_t \mid \omega_t=\omega(i), \omega_{t-1}=\omega(j)\right\}\tilde{p}_{\mbox{p}}(i, j, t)}{\frac{1}{N^2}\sum_{i=1}^N\sum_{j=1}^N p\left\{y_t \mid \omega_t=\omega(i)\right\}p\left\{z_t \mid \omega_t=\omega(i), \omega_{t-1}=\omega(j)\right\}\tilde{p}_{\mbox{p}}(i, j, t)} \label{eqn:NGF}
\end{align}
\begin{equation}
  \tilde{p}_{\mbox{s}}(i, j, t) 
  = \tilde{p}_{\mbox{f}}(i, j, t)\tilde{p}_{\mbox{s}}'(i, t+1) / \tilde{p}_{\mbox{f}}'(i, t),
\end{equation}
\noindent{where} $\tilde{p}_{\mbox{f}}'(i, t)$ and $\tilde{p}_{\mbox{s}}'(i, t+1)$ are
the marginalized filtering and smoothed distributions, respectively,
which are obtained as 
$\tilde{p}_{\mbox{f}}'(i, t) = \sum_{k=1}^N \tilde{p}_{\mbox{f}}(i, k, t) / N$
and 
$\tilde{p}_{\mbox{s}}'(i, t+1) = \sum_{k=1}^N \tilde{p}_{\mbox{s}}(k, i, t+1) / N$,
respectively.
Note in practice that the predictive and smoothed distributions should be
normalized so that the value of the integral over the whole interval becomes 1
\citep{Kitagawa2010}.

The denominator of Equation \ref{eqn:NGF} provides the approximate likelihood
for an observation at time $t$. Therefore, the log-likelihood of the model
is approximated as

\begin{align}
  &l(\boldsymbol{\xi}; Y_T, Z_T) \notag \\
  &= \sum_{t=1}^T \log \left[ \sum_{i=1}^N\sum_{j=1}^N p\left\{y_t \mid \omega_t=\omega(i)\right\}p\left\{z_t \mid \omega_t=\omega(i), \omega_{t-1}=\omega(j)\right\}\tilde{p}_{\mbox{p}}(i, j, t-1) \right]
  -2T\log N.
\end{align}
Finally, the marginal distribution for the day of initiation of menstruation
is approximated as
%
\begin{equation}
  h(k \mid Y_t, Z_t) = \frac{1}{N} \sum_{i=1}^{N} f\left\{k \mid \omega_t = \omega(i)\right\}\tilde{p}_{\mbox{f}}'(i, t).
\end{equation}
%


\section{Additional tables and figures}
\label{sect:appb}

\clearpage
\newpage

\begin{table*}[b]
  \centering
  \caption{Root mean square error (RMSE) of the sequential prediction of the next day of onset of menstruation.
    The lowest value for each subject is underlined.}
  {\footnotesize
  \begin{tabular}{lrrrrrrrrrr} \toprule
    & \multicolumn{10}{c}{Subject} \\
    Prediction time-point & 1 & 2 & 3 & 4 & 5 & 6 & 7 & 8 & 9 & 10 \\
    \midrule
    Menstrual day       & 5.418             & 4.337             & 7.536             & 2.735             & 7.738             & 4.660             & 4.273             & \underline{4.394} & \underline{2.264} & 6.909 \\             
    21 days before onset  & 7.296             & 4.235             & 9.240             & 2.993             & 8.676             & 4.175             & 4.776             & 5.820             & 10.618            & 10.817\\             
    14 days before onset  & 6.024             & 4.330             & 6.241             & 2.600             & 8.065             & 3.960             & 4.550             & 7.746             & 6.919             & 6.768 \\             
    7 days before onset   & 3.464             & 3.579             & 4.218             & 3.040             & 3.557             & 1.899             & \underline{2.487} & 5.529             & 3.062             & 3.578 \\             
    6 days before onset   & 3.144             & \underline{3.536} & 3.561             & 2.661             & 3.014             & 1.753             & 2.701             & 5.985             & 5.953             & 3.327 \\             
    5 days before onset   & 2.635             & 3.742             & 2.734             & 2.546             & 2.537             & 1.648             & 3.025             & 6.029             & 5.958             & \underline{2.160} \\ 
    4 days before onset   & 2.072             & 3.691             & 1.762             & 2.425             & 1.989             & 1.402             & 2.742             & 6.000             & 6.114             & 2.817 \\             
    3 days before onset   & \underline{2.000} & 3.691             & 0.973             & 2.366             & \underline{1.629} & 1.180             & 2.611             & 6.372             & 6.325             & 2.933 \\             
    2 days before onset   & 2.590             & 3.562             & \underline{0.795} & 2.117             & 1.830             & 1.086             & 3.097             & 6.266             & 6.666             & 3.044 \\             
    1 day before onset   & 4.022             & \underline{3.536} & 0.827             & \underline{0.663} & 1.642             & \underline{0.655} & 3.049             & 5.726             & 6.652             & 4.575 \\             
    \bottomrule
  \end{tabular}
  }
  \label{table:rmse_seq}
\end{table*}


\begin{table*}[b]
  \centering
  \caption{Root mean square error (RMSE) of the static prediction of the next day of onset of menstruation.
    The lowest value for each subject is underlined.}
  {\footnotesize
  \begin{tabular}{lrrrrrrrrrr} \toprule
    & \multicolumn{10}{c}{Subject} \\
    Predicted cycle length & 1 & 2 & 3 & 4 & 5 & 6 & 7 & 8 & 9 & 10 \\
    \midrule
    27 & 9.947 & 6.694 & 10.501 & 5.607 & 6.403  & 3.111 & 3.339 & 4.578 & 6.098 & 10.334 \\
    28 & 9.046 & 5.932 & 9.776  & 4.746 & \underline{6.338}  & 2.598 & 2.762 & 4.005 & 5.130 & 9.497  \\
    29 & 8.167 & 5.250 & 9.105  & 3.950 & 6.430  & \underline{2.413} & \underline{2.472} & 3.624 & 4.176 & 8.695  \\
    30 & 7.320 & 4.684 & 8.498  & 3.268 & 6.672  & 2.625 & 2.568 & \underline{3.495} & 3.250 & 7.937  \\
    31 & 6.517 & 4.279 & 7.970  & 2.786 & 7.050  & 3.157 & 3.012 & 3.648 & 2.385 & 7.239  \\
    32 & 5.775 & \underline{4.085} & 7.539  & \underline{2.615} & 7.541  & 3.878 & 3.682 & 4.049 & 1.677 & 6.618  \\
    33 & 5.122 & 4.131 & 7.222  & 2.814 & 8.127  & 4.702 & 4.476 & 4.634 & \underline{1.392} & 6.099  \\
    34 & 4.595 & 4.409 & 7.034  & 3.317 & 8.787  & 5.584 & 5.340 & 5.345 & 1.750 & 5.710  \\
    35 & 4.243 & 4.880 & \underline{6.985}  & 4.010 & 9.507  & 6.500 & 6.245 & 6.136 & 2.487 & 5.477  \\
    36 & \underline{4.109} & 5.494 & 7.079  & 4.812 & 10.274 & 7.438 & 7.175 & 6.981 & 3.363 & \underline{5.422}  \\
    37 & 4.215 & 6.210 & 7.309  & 5.678 & 11.079 & 8.390 & 8.122 & 7.863 & 4.294 & 5.550  \\
    \bottomrule
  \end{tabular}
  }
  \label{table:rmse_stat}
\end{table*}


\begin{table*}[b]
  \centering
  \caption{Mean absolute error (MAE) of the sequential prediction of the next day of onset of menstruation.
    The lowest value for each subject is underlined.}
    {\footnotesize
  \begin{tabular}{lrrrrrrrrrr} \toprule
    & \multicolumn{10}{c}{Subject} \\
    Prediction time-point & 1 & 2 & 3 & 4 & 5 & 6 & 7 & 8 & 9 & 10 \\
    \midrule
Menstrual day      & 6.118             & 2.875             & 4.789             & 3.000             & 3.783             & 2.286             & 2.148             & 3.826             & \underline{1.625} & 6.800  \\
21 days before onset & 7.824             & 3.563             & 7.105             & 3.040             & 3.682             & 2.357             & 2.778             & 3.957             & 10.188            & 8.600  \\
14 days before onset& 5.824             & 3.813             & 5.263             & 1.280             & 4.348             & 2.214             & 2.926             & 5.957             & 8.250             & 5.733  \\
7 days before onset& 4.000             & 2.875             & 4.579             & 2.320             & 3.826             & 1.357             & \underline{1.778} & 3.696             & 3.813             & 3.600  \\
6 days before onset& 3.353             & 2.875             & 3.842             & 1.880             & 3.217             & 1.357             & 2.074             & 3.565             & 4.250             & 3.267  \\
5 days before onset& 2.882             & 2.938             & 3.105             & 1.760             & 2.957             & 1.214             & 2.259             & 3.304             & 3.688             & 2.000  \\
4 days before onset& 2.294             & 2.813             & 2.000             & 1.520             & 2.087             & 0.893             & 2.111             & 3.000             & 3.063             & 2.067  \\
3 days before onset& 1.765             & 2.375             & 1.105             & 1.360             & 1.043             & 0.821             & 1.815             & 3.000             & 2.250             & \underline{1.333}  \\
2 days before onset& \underline{1.294} & 2.000             & 0.368             & 1.080             & 0.783             & 0.714             & 1.926             & 2.304             & 2.375             & \underline{1.333}  \\
1 day before onset& 1.765             & \underline{1.813} & \underline{0.211} & \underline{0.280} & \underline{0.348} & \underline{0.321} & \underline{1.778} & \underline{0.739} & 1.875             & 1.467  \\
    \bottomrule
  \end{tabular}
  }
  \label{table:mae_seq}
\end{table*}


\begin{table*}[b]
  \centering
  \caption{Mean absolute error (MAE) of the static prediction of the next day of onset of menstruation.
    The lowest value for each subject is underlined.}
  {\footnotesize
  \begin{tabular}{lrrrrrrrrrr} \toprule
    & \multicolumn{10}{c}{Subject} \\
    Predicted cycle length & 1 & 2 & 3 & 4 & 5 & 6 & 7 & 8 & 9 & 10 \\
    \midrule
    27 & 9.059             & 5.313             & 7.947             & 4.960             & \underline{3.870} & 2.179             & 2.778             & 3.130             & 5.938             & 8.800 \\
    28 & 8.059             & 4.313             & 7.053             & 3.960             & 4.174             & \underline{1.750} & 2.222             & 2.826             & 4.938             & 7.933 \\
    29 & 7.176             & 3.688             & 6.158             & 3.040             & 4.739             & 1.893             & 1.889             & \underline{2.783} & 3.938             & 7.200 \\
    30 & 6.294             & 3.188             & 5.263             & 2.440             & 5.391             & 2.250             & \underline{1.852} & 2.826             & 2.938             & 6.600 \\
    31 & 5.412             & 3.063             & 4.368             & 2.080             & 6.130             & 2.821             & 2.333             & 3.130             & 1.938             & 6.000 \\
    32 & 4.529             & \underline{2.938} & 4.000             & \underline{2.040} & 6.870             & 3.464             & 3.111             & 3.522             & 1.313             & 5.400 \\
    33 & 3.765             & \underline{2.938} & \underline{3.737} & 2.400             & 7.609             & 4.321             & 3.963             & 4.087             & \underline{1.188} & 4.800 \\
    34 & 3.118             & 3.313             & 3.789             & 2.840             & 8.348             & 5.179             & 4.889             & 4.826             & 1.438             & 4.200 \\
    35 & \underline{2.824} & 3.938             & 4.158             & 3.520             & 9.087             & 6.107             & 5.815             & 5.565             & 2.063             & \underline{3.733} \\
    36 & 3.000             & 4.813             & 4.737             & 4.280             & 9.826             & 7.036             & 6.741             & 6.391             & 3.063             & 3.800 \\
    37 & 3.412             & 5.688             & 5.421             & 5.200             & 10.565            & 8.036             & 7.741             & 7.217             & 4.063             & 4.133 \\
    \bottomrule
  \end{tabular}
  }
  \label{table:mae_stat}
\end{table*}


\begin{figure*}[tbp]
{\centering
\includegraphics[bb=0 0 864 648,width=6in]{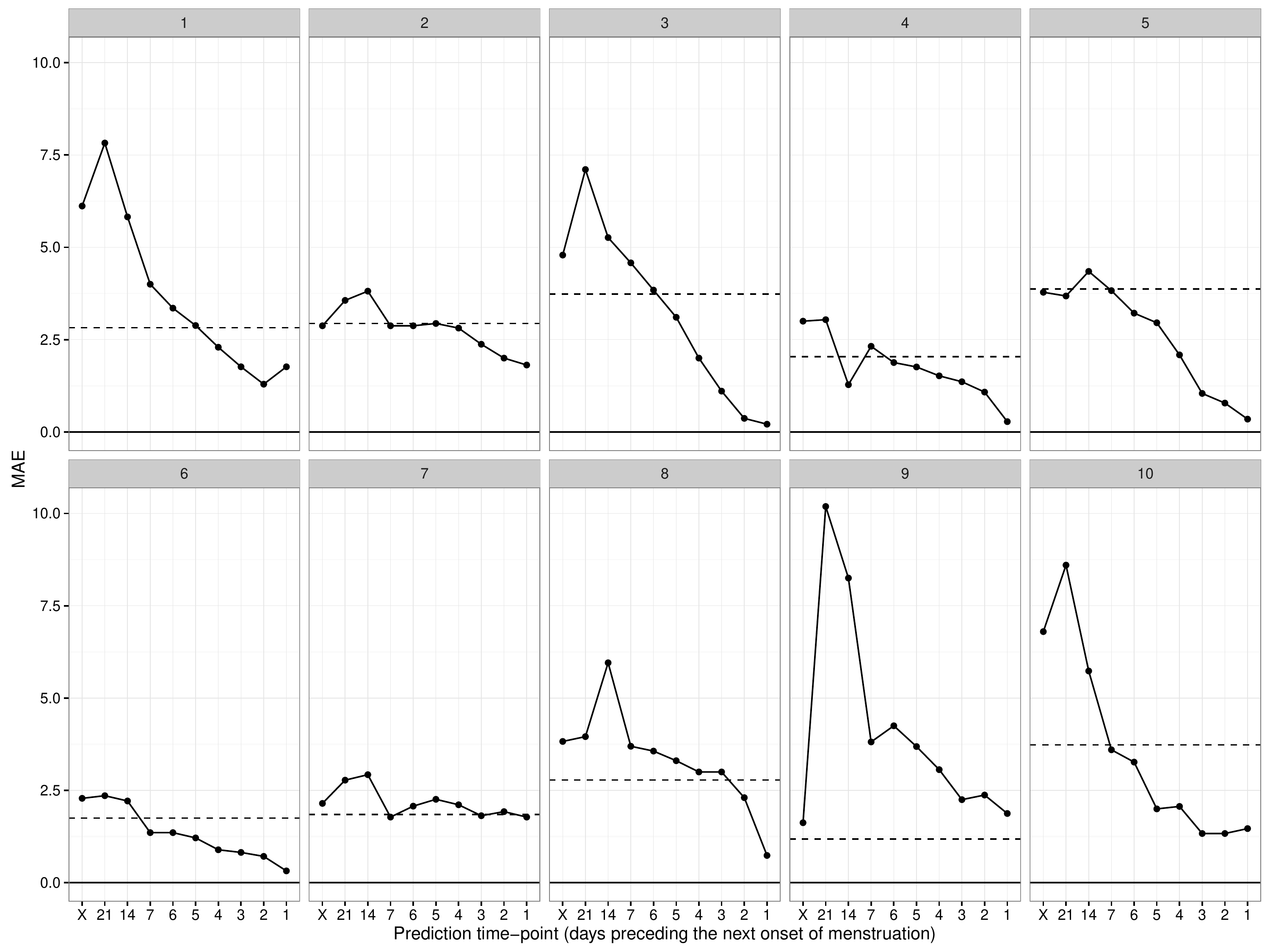}
}
\caption{
  Mean absolute error (MAE) of the prediction of the day of onset of the next menstruation.
  Solid lines indicate the MAE of the sequential prediction
  (based on the best AIC model for each subject),
  and horizontal dashed lines indicate the lowest MAE of the conventional
  prediction for each subject.
  On the horizontal axis, ``X'' indicates the prediction obtained at the day
  of onset of the previous menstruation.
  Each panel shows the results for one subject.
  Compared with the best prediction provided by the conventional method, 
  the range, mean and median of the rate of the maximum reduction of MAE 
  for each subject by using the sequential method was 
  0.056--1.368, 0.449 and 0.311, respectively.
}
\label{fig:mae}
\end{figure*}

\end{document}